\documentclass[10pt,a4paper]{iopart}
\usepackage[left=2.3cm, top=4.5cm]{geometry} %manually correct offsets
\usepackage[english]{babel} %neu
\usepackage[hidelinks]{hyperref}
\usepackage{iopams}  
\usepackage{tabularx}
\usepackage{mdframed}
\usepackage{orcidlink}
\usepackage{tikz}
\usepackage{pgfplots}
\pgfplotsset{compat=1.18} 
\usetikzlibrary{positioning, calc, decorations.pathreplacing}
\usetikzlibrary{intersections, positioning}
\usetikzlibrary{calc,decorations.pathmorphing}
            
\usepackage{blindtext}
\usepackage[nolist,nohyperlinks]{acronym}

\begin{acronym}[ECU]

\acro{MMR}[MMR]{magneto-mechanical resonator}
\acro{Rx}[Rx]{receive window}
\acro{SNR}[SNR]{signal-to-noise ratio}
\acro{Tx}[Tx]{transmit window}

\end{acronym}

\usepackage{graphicx, xcolor}
\graphicspath{{./graphics/}}

\usepackage[exponent-product = \cdot]{siunitx}
\DeclareSIUnit{\mup}{\text{$\mu_0$}}
\DeclareSIUnit{\mT}{\milli\tesla}
\DeclareSIUnit{\Tpm}{\tesla\per\metre}
\DeclareSIUnit{\mul}{\micro\litre}
\DeclareSIUnit{\degC}{\celsius}    
\DeclareSIUnit{\mgfeml}{\milli\gram\of{Fe}\per\milli\liter}

\usepackage[style=ieee,
            citestyle=numeric-comp, 
            url=false,doi=true,
            maxbibnames=99,
            ]{biblatex} 

\begin{document}

\title{Wireless and passive pressure detection using magneto-mechanical resonances in process engineering}

\author{Timo Merbach\textsuperscript{1}\,\orcidlink{0000-0002-7723-5444}, 
Felix Kexel\textsuperscript{1}\,\orcidlink{0000-0003-4268-2348}, 
Jonas Faltinath\textsuperscript{2,3}\,\orcidlink{0009-0003-4128-2948}, 
Martin M\"oddel\textsuperscript{2,3}\,\orcidlink{0000-0002-4737-7863},
Michael Schlüter\textsuperscript{1}\,\orcidlink{0000-0001-5969-2150},
Tobias Knopp\textsuperscript{2,3,4}\,\orcidlink{0000-0002-1589-8517}, and
Fabian Mohn\textsuperscript{2,3}\,\orcidlink{0000-0002-9151-9929}}

\address{\tiny\textsuperscript{1}\scriptsize Institute of Multiphase Flows, Hamburg University of Technology, Hamburg, Germany}
\address{\tiny\textsuperscript{2}\scriptsize Institute of Biomedical Imaging, Hamburg University of Technology, Hamburg, Germany}
\address{\tiny\textsuperscript{3}\scriptsize Section of Biomedical Imaging, Medical Center Hamburg-Eppendorf, Hamburg, Germany}
\address{\tiny\textsuperscript{4}\scriptsize Fraunhofer Research Institution for Individualized and Cell-based Medical Engineering IMTE, L\"ubeck, Germany}

\ead{\footnotesize \href{mailto:timo.merbach@tuhh.de}{timo.merbach@tuhh.de} \hfill February 2025}

\begin{abstract}
%Info by IOP: single column, max. 300 words. The abstract should be complete in itself; it should not contain undefined acronyms/abbreviations and no table numbers, figure numbers, references or equations should be referred to.
A custom-developed magneto-mechanical resonator (MMR) for wireless pressure measurement is investigated for potential applications in process engineering. The MMR sensor utilises changes in the resonance frequency caused by pressure on a flexible 3D printed membrane. The thickness of the printed membrane plays a crucial role in determining the performance and sensitivity of MMRs, and can be tailored to meet the requirements of specific applications. The study includes static and dynamic measurements to determine the pressure sensitivity and temporal resolution of the sensor. The results show a minimum sensitivity of $0.06~\text{Hz mbar}^{-1}$ and are in agreement with theoretical calculations and measurements. The maximum sensor readout frequency is 2~Hz in this study. Additionally, the temperature dependence of the sensor is investigated, revealing a significant dependence of the resonance frequency on temperature. The developed MMR offers a promising and versatile method for precise pressure measurements in process engineering environments.
\end{abstract}

%
% Uncomment for keywords
\vspace{2pc}

\ioptwocol

% Uncomment for Submitted to journal title message
%\submitto{\JPA}

%%%%%%%%%%%%%%%%%%%%%%%%%%%%%%%%%%%%%%%%%%%%%%%%%%%%%%%%%%
\section{Introduction}
In process engineering, hydrodynamic process data, such as residence times, mixing times, and volume flow exchange rates, are crucial for scale-up, process control, and optimisation~\cite{siebler2019impact, vrabel2000mixing, rosseburg2018hydrodynamic}. In both industrial and academic settings, conventional sensors are typically mounted on the exterior of the vessel and rely on a Eulerian approach for tracking. In contrast, Lagrangian sensors are deployed directly within the moving fluid inside the vessel, providing insights into the internal process conditions. Numerous research groups are actively developing such sensors, namely Lagrangian sensor particles, particularly for the application in bioreactors~\cite{buntkiel2023orientation, hofmann2024experimental, bisgaard2021characterization, rautenbach2024dynamics, stine2020electrochemical}. A significant challenge with these sensors lies in determining their position within the vessel. Currently, only axial sensor data—obtained via optical access or hydrostatic pressure—can be utilised for this purpose~\cite{hofmann2024experimental, bisgaard2021characterization, rautenbach2024dynamics}. Furthermore, energy supply of active sensors is challenging~\cite{lauterbach2019measurements}. \par
The medical field is also confronted with the challenge of monitoring the position and orientation of medical instruments within the human body, necessitating the acquisition of radial and axial information. In response to this need, a novel wireless and passive sensor based on magneto-mechanical resonance was recently introduced~\cite{gleich_miniture_2023}. While this sensor is focused on miniaturisation for medical applications such as endoscopy, surgery, implantation, and vascular interventions to measure position and orientation (6~degrees of freedom), it also has huge potential for technical applications such as process engineering for which is has not yet been utilised so far. In short, the so-called magneto-mechanical resonator~(MMR) can be used for tracking by measuring its directional response to a weak magnetic field excitation. In addition, the MMR allows for sensing, since the internal magnet distance affects its resonance frequency \cite{knopp_empirical_2024}. Coupling this distance to an additional mechanical sensing mechanism allows for determining an environmental parameter such as temperature, pressure, magnetic fields or viscosity.  \par
One crucial aspect of data analysis is the possibility of co-registering wireless sensor data with spatial tracking information. As demonstrated in~\cite{gleich_miniture_2023}, spatial tracking with MMRs has been successfully achieved. However, the integration of tracking and sensing processes introduces additional challenges in signal processing that are beyond the scope of the present study. Instead, the study is focused solely on investigating the sensing capabilities of the MMR sensor in the context of process engineering as a proof of concept. Such applications present a number of additional challenges, including electromagnetic interference, distorted supply currents, non-transparent stainless-steel vessels, and extreme temperature as well as pressure conditions. Consequently, this study develops and tests an MMR sensor for measuring pressure in a liquid column under a range of such conditions. It focuses on the sensor's sensitivity, measurement range, real-time accuracy, and temperature cross-sensitivity. Three experimental approaches are used: static experiments with constant pressure, dynamic experiments measuring pressure variation, and temperature-changing MMR tests.

\begin{figure}[b!]
    \centering
    \includegraphics[width=1.0\linewidth]{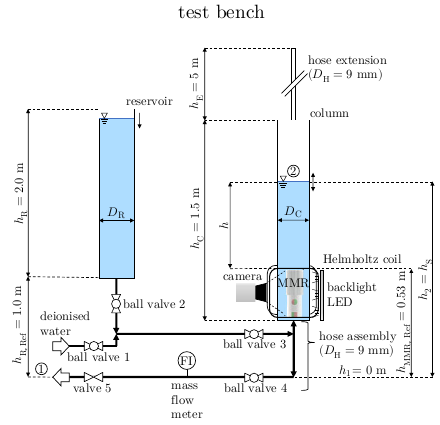}
    \caption{\textbf{Test bench for hydrostatic and hydrodynamic experiments.} Schematic flow diagram of the test bench detailing the column, reservoir, and associated peripheral systems.}
    \label{fig:Flow-diagram-drawing}
\end{figure}

%%%%%%%%%%%%%%%%%%%%%%%%%%%%%%%%%%%%%%%%%%%%%%%%%%%%%%%%%
\section{Material and methods}
A refillable acrylic column is employed to establish different pressure levels on the MMR, while the passive natural resonance of the MMR is excited inductively and received using a set of coils, located on the outside of the column. The following sections will provide a detailed explanation of each component of the setup.

\subsection{Hydrodynamic setup}
Figure~\ref{fig:Flow-diagram-drawing} shows the schematic flow diagram of the test bench. A transparent cylindrical column with an inner diameter $D_\text{C} = 70~\text{mm}$ is employed for the experiments. The column has a height of $h_\text{C} = 1.5$~m, with an attached hose extending the column to a total of 6~m. This setup follows the Pascal's paradox, where the pressure depends solely on the height of the column, regardless of the tank's volume or shape~\cite{pascal1663traitez, spurk2007fluid}. A measuring tape is affixed externally along the full length to facilitate optical height measurement (with an uncertainty of $\pm 1$~cm), thereby enabling subsequent hydrostatic pressure determination. Additionally, due to the optical access provided by the column and the MMR, a camera using backlight imaging is employed to analyse the edge-to-edge distance~$d_\text{e}$ between the magnets during the experiment, thus offering further validation of the measurement principle. \par
\begin{figure*}[t!]
    \centering
    \includegraphics[width=1.0\linewidth]{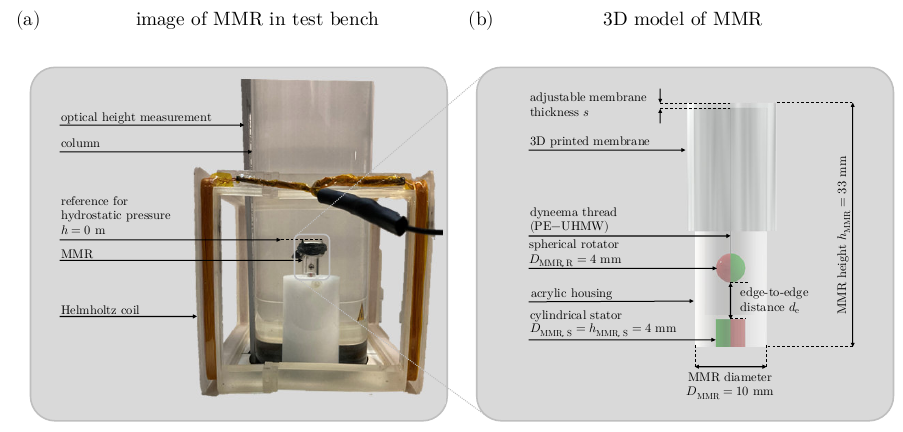}
    \caption{\textbf{MMR setup and implementation in the test bench.} In (a), MMR in the test bench integrated with the square-shaped Helmholtz coil and fixed within the column for the purpose of measuring the hydrostatic pressure. 3D model of the MMR showcasing the 3D printed membrane, rotator, and stator in (b).}
    \label{fig:MMR-drawing}
\end{figure*}

 
The column is filled with deionised water maintained at a constant temperature~$T = (21.1 \pm 0.5)\ ^\circ\text{C}$, with a corresponding fluid density~$\rho = (998.0 \pm 0.1)\ \text{kg m}^{-3}$~\cite{stephan2019vdi}. A reservoir ($D_\text{R}=0.29~\si{\metre}, h_\text{R}=2.0~\si{\metre}$) is connected to the column in order to supply the latter with deionised water, with the two components connected by the principle of communicating liquid levels. The deionised water is conditioned to ambient temperature over several hours in the reservoir of the test bench, ensuring thermal equilibrium between deionised water and surrounding air. This preparation eliminates thermal effects in the column and MMR, preventing temperature-induced variations in the results. In some static experiments, unconditioned deionised water is used, given the reservoir height is below the maximum height of the test bench. The added volume of the unconditioned water is minimal, as the column is already filled, and the inner diameters of the hoses located above it are significantly smaller ($D_\text{C}\gg D_\text{H}=9~\si{\mm}$). Additionally, a waiting period allows temperature differences to equilibrate. The MMR is mounted to the base of a column at a height $h_\text{MMR, Ref} = 0.53$~m above the outlet. Figure~\ref{fig:MMR-drawing}~(a) shows the configuration surrounding the MMR, depicting its arrangement in the test bench. \par
Dynamic experiments will be conducted, where the inlet and outlet dynamics are characterised. In the case of the inlet dynamics, the column is fed from the reservoir, which has a significantly larger volume. For the outlet dynamics, the outlet of the column is connected to a hose assembly, located beneath the column, as shown in Figure~\ref{fig:Flow-diagram-drawing}. The hoses also have an inner diameter of $D_\text{H} = 9$~mm. The segment of the assembly corresponding to the outlet dynamics features a ball valve (ball valve~4), followed by a Coriolis mass flow meter (Promass 80, Endress + Hauser, Switzerland) for measuring the outlet flow, and an additional valve (valve~5), which is maintained in a fixed position. To discharge the column the ball valve is fully opened, resulting in a consistent and reproducible outlet time. To evaluate the inlet dynamics, the column is filled from the bottom via ball valves~2 and~3, ensuring immediate pressure build-up at the sensor for accurate and responsive tracking of liquid level changes. This way promotes a smooth, con-

\begin{mdframed} 
\subsection*{\textnormal{\textbf{Nomenclature}}}
\subsection*{\textnormal{\textbf{Arabic symbols}}}
\begin{tabularx}{\textwidth}[t!]{@{}lX@{}}
$D_\text{C}$ & Column diameter / m \\
$D_\text{H}$ & Hose diameter / m \\
$D_\text{MMR}$ & MMR diameter / m \\
$D_\text{MMR, R}$ & Rotator diameter / m \\
$D_\text{MMR, S}$ & Stator diameter / m \\
$D_\text{R}$ & Reservoir diameter / m \\
$d_0$ & Initial distance / m \\
$d_\text{e}$ & Edge-to-edge distance / m \\
$d_\text{m}$ & Mass-centred distance / m \\
$\Delta d$ & Stretch distance / m\\
$F$ & Force / N \\
$f_\text{nat}$ & Natural frequency / Hz \\
$g$ & Gravitational acceleration  / m s$^{-2}$ \\
$h$ & Height for hydrostatic pressure  / m \\
$h_\text{1}$ & Height at position \raisebox{.5pt}{\textcircled{\raisebox{-.9pt} {1}}}  / m \\
$h_\text{2}$ & Liquid level at position \raisebox{.5pt}{\textcircled{\raisebox{-.9pt} {2}}}  / m \\
$h_\text{C}$ & Column height  / m \\
$h_\text{E}$ & Height of hose extension  / m \\
$h_\text{MMR}$ & MMR height  / m \\
$h_\text{MMR, S}$ & Stator height  / m \\
$h_\text{R}$ & Reservoir height  / m \\
$h_\text{R, Ref}$ & Reference height for reservoir  / m \\
$h_\text{MMR, Ref}$ & Reference height for MMR  / m \\
$h_\text{S}$ & Height for outlet dynamics  / m \\
$I$ & Inertia  / kg\,m$^2$\\
$L$ & Inductance  / H \\
$m_\text{R}$ & Rotator magnetic moments  / A m$^2$ \\
$m_\text{S}$ & Stator magnetic moments  / A m$^2$ \\
$n$ & Number of measurements  / $-$\\
$p$ & Pressure / Pa \\
$p_0$ & Atmospheric pressure  / Pa \\
$p_\text{1}$ & Pressure at position \raisebox{.5pt}{\textcircled{\raisebox{-.9pt} {1}}}   / Pa \\
$p_\text{2}$ & Pressure at position \raisebox{.5pt}{\textcircled{\raisebox{-.9pt} {2}}}  / Pa \\
$R$ & Resistance  / $\Omega$ \\
$r^2$ & Coefficient of determination  / $-$ \\
$s$ & Membrane thickness  / m \\
$T$ & Temperature  / $^\circ $C \\
$T_0$ & Ambient temperature  / $^\circ $C \\
$t$ & Time  / s \\
$t_\text{rx}$ & Receive time  / s\\
$t_\text{tx}$ & Transmit time  / s\\
$\dot V$ & Volumetric flow rate / m$^3$ s$^{-1}$ \\
$v_\text{1}$ & Liquid velocity at position \raisebox{.5pt}{\textcircled{\raisebox{-.9pt} {1}}}  / m s$^{-1}$\\
$v_\text{2}$ & Liquid velocity at position \raisebox{.5pt}{\textcircled{\raisebox{-.9pt} {2}}}  / m s$^{-1}$\\
\end{tabularx}

\subsection*{\textnormal{\textbf{Greek symbols}}}
\begin{tabularx}{\textwidth}{@{}lX@{}}
$\zeta$ & Resistance coefficient / $-$ \\
$\mu_0\ \ \ \ \ \ \ \ \ \  $ & Vacuum permeability  / H m$^{-1}$ \\
$\rho$ & Fluid density  / kg m$^{-3}$ \\
\end{tabularx}
\end{mdframed} 

\newpage
\noindent
tinuous rise in liquid without significant surface turbulence. The inlet and outlet dynamics are based on established hydrodynamic principles, as optical level measurement of rapidly changing liquid levels is challenging. Only the upper and lower levels are determined optically, while the dynamics are assessed by recording the time taken to fill or discharge a 1~metre liquid column. 

\subsection{Hydrostatic and hydrodynamic fundamentals}
The hydrostatic pressure~$p(h)$ on the MMR is directly proportional to the height of the liquid level above the MMR~$h$ as described in Equation~(\ref{eq:Hydrostatic}). The reference height for the hydrostatic pressure is defined at the top of the MMR membrane, as shown in Figure~\ref{fig:MMR-drawing}~(a). For a fluid with constant density~$\rho$, the hydrostatic pressure is given by Pascal's law as
\begin{equation}
    \label{eq:Hydrostatic}
    p(h)-p_0 = \rho g h, 
\end{equation}
where $p_0=1$~atm denotes the atmospheric pressure and $g=9.81\ \text{m s}^{-2}$ the gravitational acceleration~\cite{pascal1663traitez}. All pressures are given as relative pressures in this paper. \par 
In addition, dynamic experiments are also carried out. The inlet and outlet dynamics are characterised mathematically, as detailed below. For the inlet dynamics, the volumetric flow rate~$\dot V$ is assumed to be constant. This assumption is based on the fact that 
the test bench reservoir is considerably larger than the column and has a significantly higher liquid level. Therefore, 
\begin{equation}
    \label{eq:Const-flow}
    h(t) = \frac{4\dot V t}{\pi D_\text{C}^2}
\end{equation}
can describe the inlet dynamics, which represent the change in height and hydrostatic pressure as a function of time~$t$~\cite{spurk2007fluid}. The Bernoulli equation 
\begin{equation}
    \label{eq:Bernoulli}
  \frac{v_1^2\rho}{2}+\rho g h_1 + p_1 +  \frac{\zeta v_1^2 \rho}{2} =  \frac{v_2^2\rho}{2}+\rho g h_2 + p_2
\end{equation}
is essential for analysing the outlet dynamics, as it considers for variations in liquid height. Equation~(\ref{eq:Bernoulli}) accounts for frictional flow caused by the components within the outlet section, with $\zeta$ representing the total resistance coefficient and $v_\text{i}$ denotes the velocity in the column or the outlet hose~\cite{spurk2007fluid}. The indices refer to the positions relevant for the Bernoulli equation, namely the outlet hose (\raisebox{.5pt}{\textcircled{\raisebox{-.9pt} {1}}}) and the liquid level in the column (\raisebox{.5pt}{\textcircled{\raisebox{-.9pt} {2}}}), as illustrated in Figure~\ref{fig:Flow-diagram-drawing}. \par
In deriving the analytical solution of Equation~(\ref{eq:Bernoulli}), assumptions such as $p_\text{1}=p_\text{2}=p_0$, $v_\text{2}\approx0$~m~s$^{-1}$, $h_\text{1}=0$~m, and $\rho = \text{const.}$ are made, from which the modified Torricelli equation
\begin{equation}
    \label{eq:modified-Torri}
   v_\text{1}=\sqrt{\frac{2gh_\text{2}}{1+\zeta}}
\end{equation}
is obtained. \newpage
Equation~(\ref{eq:modified-Torri}) is used in conjunction with the mass balance 
\begin{equation}
    \label{eq:mass-balance}
   -D^2_\text{C}\frac{dh_\text{2}}{dt} = v_\text{1} D^2_\text{H}
\end{equation}
for the column~\cite{spurk2007fluid}. The integration from $h_\text{S}$ to $h(t)$ and $0$ to $t$ yields in
\begin{equation}
    \label{eq:Toricelli}
    h(t) = \left (\sqrt{h_\text{S}}-t\left(\frac{D_\text{H}}{D_\text{C}}\right)^2 \sqrt{\frac{g}{2(1+\zeta)}} \right )^2
\end{equation}
as a means of characterising the dynamics of the outlet. $h_\text{S}$ is the set starting height in the column above the outlet hose. Based on the known run-out time and rearranged Equation~(\ref{eq:Toricelli}) the total resistance coefficient~$\zeta$ is determined at $\zeta = 19.9 \pm 0.7$. The calculated resistance coefficients are compared and verified with the values found in the literature and those provided by the manufacturer~\cite{spurk2007fluid, idelchik1986handbook, Endress}. Moreover, a mass flow meter is employed to measure the mass flow rate at the outlet, enabling experimental validation of the outlet curve. The conversion of the mass flow rate into height as a function of time is accomplished via numerical integration of the measured values over time, incorporating the water's density and the column's cross-sectional area.

\subsection{Magneto-mechanical resonator (MMR)}
\label{sec:methods:mmr}
Figure~\ref{fig:MMR-drawing}~(b) shows the MMR developed for pressure measurement applications. At the core of the sensor are two permanent magnets, separated by a distance, and a flexible membrane. One permanent magnet is a spherical neodymium magnet (N40) with a diameter  $D_\text{MMR, R}=4$~mm, referred to as the rotator. The rotator is connected to the membrane via a thin Dyneema thread (PE-UHMW), fixed to the magnet with epoxy resin. The stator, a diametral magnetised cylindrical neodymium magnet (N35) with a height and diameter $D_\text{MMR, S}=h_\text{MMR, S}=4$~mm, is fixed and embedded within the cylindrical acrylic housing. The housing has a height $h_\text{MMR}=33$~mm and features a base diameter $D_\text{MMR}=10$~mm. The working principle of the MMR is based on the antiparallel arrangement of the stator and the rotator~\cite{gleich_miniture_2023}, which attract each other with a force several orders of magnitude greater than gravity and with the freedom for torsional oscillation.
Following excitation by external magnetic fields, the rotator exhibits a damped oscillatory dynamics converging to its natural frequency $f_\text{nat}$ with a high quality factor. Analogue signal processing hardware is used to switch between transmit (Tx) and receive (Rx) windows with times $t_\text{tx}$ (transmit) and $t_\text{rx}$ (receive), thus controlling the oscillation in real time and allowing $f_\text{nat}$ to be read wirelessly. One possible way to implement a sensor is by changing the mass-centred distance~$d_\text{m}$ between both magnet mass-centres to cause a desired change in frequency. This is done by fixing the rotator to a flexible membrane and sealing the MMR housing. The dependence of MMR resonance on frequency can be approximated by the dipole far field model for spheres~\cite{knopp_empirical_2024} with
\begin{equation}\label{eq:fnat}
    f_\text{nat}(d_\text{m}) = \frac{1}{2 \pi} \sqrt{\frac{m_\text{R} m_\text{S} \mu_0}{4\pi d_\text{m}^3 I}} \propto \, d_\text{m}^{-\frac{3}{2}} \; .
\end{equation}
The magnetic moments of the rotator and the stator $m_\text{R}$, $m_\text{S}$, vacuum permeability $\mu_0$, mass-centred distance $d_\text{m}$, and the inertia $I$ determine the resonance frequency. In this instance and within the measurement range, these physical parameters are assumed to be constant, and the significant parameter is the distance~$d_\text{m}$ only. The initial frequency of a sensor is $f_0 = f_\text{nat}(d_0)$, where $d_0$ refers to the equilibrium distance at atmospheric pressure~$p_0$ and ambient temperature $T_0$ (sealing conditions). The mechanical range of the sensor is limited to a displacement of approximately $\Delta d \leq 6$~mm, where the magnets will come into contact in the extreme case if edge-to-edge distance between the magnets~$d_\text{e}$ approaches zero (maximum pressure). It can be observed that an initial distance exceeding $d_0 = 6$~mm would result in an insufficiently strong attraction and a markedly low resonance frequency when the separation of magnets becomes too large. 

\subsection{3D printed elastic membrane}
\label{sec:methods:membrane}
\begin{figure}[t!]
    \centering
    \includegraphics[width=1.0\linewidth]{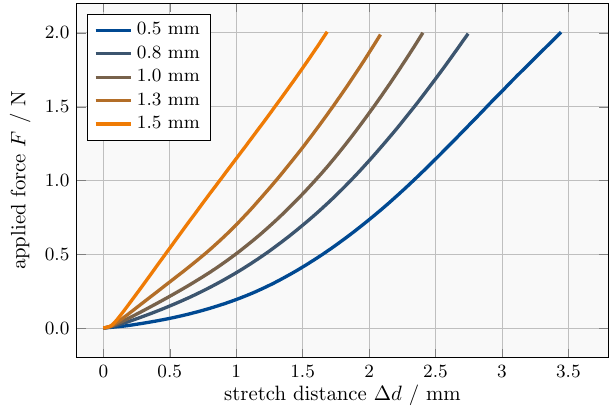}
    \caption{\textbf{Membrane design and parameter choice.} The stretch distance~$\Delta d$ in response to applied force of different membrane thicknesses between $s=0.5$~mm and $1.5$~mm thickness, 3D printed with Formlabs Elastic 50A V1 resin, are investigated with a MiniZwick testing machine with 1~mm punch.}
    \label{fig:force}
\end{figure}

The MMR is equipped with a membrane fabricated using a Form 3+ (Formlabs, USA) 3D printer through the stereolithography process. The membrane material is Elastic 50A V1 (Formlabs, USA), a flexible elastomer. The membrane is approved by the manufacturer for water resistance and has an size expansion of less than 1~\%~\cite{Formlabs}. The membrane thickness can be precisely controlled to an accuracy of 100~µm, limited by the 3D printer. The performance and sensitivity of MMRs are significantly influenced by the thickness of the printed membrane~$s$. The relationship between membrane thickness and pressure sensitivity is experimentally validated through a force test using a MiniZwick testing machine equipped with a 10~N load cell (ZwickRoell GmbH \& Co. KG, Germany) and a 1~mm diameter punch. The test involved pressing the membrane at the centre of up to a force $F=2$~N, which corresponds to the expected hydrostatic pressure in the test bench. The resulting stretch distance~$\Delta d$ is measured in the transverse direction. Figure~\ref{fig:force} illustrates the results for various membrane thicknesses in a range from $s = 0.5$~mm to  $1.5$~mm, confirming the existence of a direct correlation between thickness and sensitivity. Based on these findings, the membrane thickness must be selected based on the anticipated pressure range for the application. For this work, a membrane with a thickness $s=0.8$~mm is chosen. This thickness provides high sensitivity to small pressure variations as a trade-off between sensitivity, membrane durability, and maximum applicable pressure to the sensor, before magnets touch. 

\subsection{Inductive measurement setup}
To excite the passive MMR sensor, a separable square-shaped Helmholtz-like coil is used, that can be joined around the base of the column (see Figure~\ref{fig:MMR-drawing}~(a)), chosen for its near homogeneous field distribution~\cite{hurtado-velasco_simulation_2016}. The coils differ in shape and are twice the distance from a true Helmholtz coil setup ($L=910\,\text{\textmu H},\; R=5.1\,\Omega$). Magnetic fields in the range of 5 to $100$~\textmu T $\mu_0^{-1}$ and $100$ to $300$~Hz are used to pump the torsional oscillation to an deflection angle of about $(11 \pm 2)$\;degree during the transmit window. Based on reciprocity, the receive window is used to capture the MMR signal via the same coil pair. Transmitted signals are amplified by a class-D amplifier and received signals by a low-noise amplifier~\cite{mohn_low-cost_2025}, both controlled by the combined  RedPitaya Stemlab 125-14 DAC/ADC card operated on a custom software stack~\cite{hackelberg2022flexible}. Low-level signal processing is done in real-time using frequency and phase-controlled re-excitation of the MMR. An analogue filter is used to attenuate $50$~Hz harmonics in the power supply for a cleaner receive spectrum. 

\subsection{Sensing parameter estimation}
The natural frequency $f_\text{nat}$ can be determined by fitting the damped oscillator model to the measured data~\cite{gleich_miniture_2023}. As neither the change over distance in Equation~(\ref{eq:fnat}) nor the stretch distance $\Delta d$ of the membrane in Figure~\ref{fig:force} are linear with frequency, a calibration-based estimator is employed. The combined system is measured at a range of known static pressure levels with $20$~averages (see static experiments in Section~\ref{sec:res:static}). These values are fitted by a quadratic polynomial using the least squares method, and its inverse serves as an estimator for pressure for dynamic measurements (see Section~\ref{sec:res:dynamic}). The estimator is limited to the relevant dynamic range between 0 and $100$~mbar in this work and allows to derive the MMR sensitivity in Hz\;mbar$^{-1}$ by taking the derivative of the polynomial fit.

\begin{figure*}[t!]
    \centering
    \includegraphics[width=1.0\linewidth]{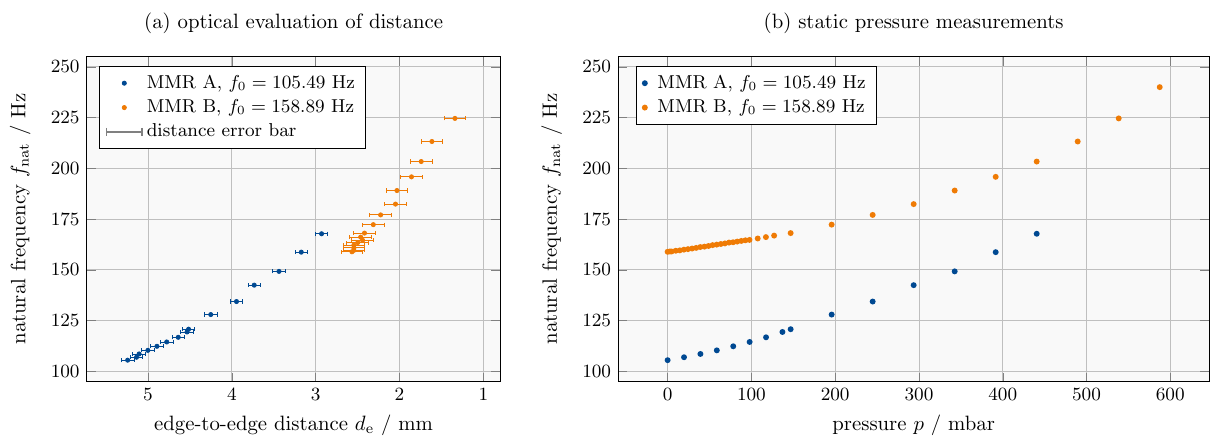}
    \caption{\textbf{Results of static MMR measurements.} The change in edge-to-edge distance~$d_\text{e}$ of the two magnets within the MMR is determined optically by inspecting images in (a). As the pressure increases, the membrane becomes convex and decreases the distance of stator and rotator magnets, which increases the natural frequency. The same measurements are plotted in (b), where measurements are mapped to the corresponding hydrostatic pressure of the column (optical observation). For MMR~B, 10 averages per point and 5\;mbar increments are used at the beginning, while for MMR~A, 20 averages and 20\;mbar increments are applied. For both MMRs, increments of 50\;mbar are used above 150\;mbar. Overall standard deviation for MMR~A and B frequencies average to $0.2$~Hz and $0.4$~Hz, respectively.}
    \label{fig:res-dist-and-static}
\end{figure*}

\subsection{Experimental procedure}
Three different experimental approaches are employed: static experiments, where the pressure is held constant over time; dynamic experiments, which aim to measure the pressure variation over time; and experiments designed to explore the temperature dependence of the MMR. Three equal MMRs (MMR~A, B, C) are used for the experiments, all based on the design shown in Figure~\ref{fig:MMR-drawing}~(b). The only differences between them are the initial distances of the magnets~$d_0$, resulting in varying initial frequencies~$f_0$. For all experiments, the MMRs are sealed at ambient temperature $T_0 = (21.1 \pm 0.5)~^\circ\text{C}$, ensuring that the membrane is in an equilibrium state at $p_0$ and $T_0$. Additionally, all MMRs are inserted in water prior to experiments to account for temperature equilibrium throughout the sensor and for minuscule membrane water uptake (see Section~\ref{sec:methods:membrane}). \par
The \textbf{static experiments} use a stepwise increase in pressure, where measurements are taken at constant pressure levels with at least $10$~frame averages at each level. These experiments are also captured on camera to optically evaluate the changing edge-to-edge distance and are repeated for the MMR~A and B. Image-based analysis are carried out in \textit{MATLAB} (MathWorks Inc., USA), with calibration based on the known diameter of the rotator. A TxRx sequence with a long readout time is used to ensure that the exponential decay of the signal is captured to reduce errors (sequence cycle: $t_\text{tx}=50\dots 100$\,ms, $t_\text{rx}=2000\dots 4000$\,ms). The sensitivity in $\text{Hz mbar}^{-1}$ is obtained by using linear regression in the lower sensor region between $0$ and $100$~mbar. \par

To capture \textbf{dynamic changes}, the liquid level is raised from 0 to $\approx 1$~m in 33~s, held for 23~s, and then discharged to 0~m over 63~s. The timing is measured manually using a stopwatch (with an uncertainty of $\pm 1$~s), and the height of the column is adjusted manually using ball valves. During this experiment, a 2~Hz frame rate is used (sequence cycle: $t_\text{tx}=100$\,ms, $t_\text{rx}=400$\,ms) to measure MMR~A in real time. As explained earlier a mass flow meter is used as an additional reference during the discharge.

In contrast to the previous experiments, where the temperature is kept constant, the third experiment specifically investigates the \textbf{temperature dependence} of the MMRs. In this experiment, MMR~C is submerged in a water tank at a constant depth, ensuring a constant hydrostatic pressure of 15~mbar, which is maintained throughout the experiment. Since the density of water affects the hydrostatic pressure and is temperature-dependent, temperature variations also influence the hydrostatic pressure. However, within the examined temperature range, the change in density is less than 2~\% and can therefore be regarded as negligible~\cite{stephan2019vdi}. Temperature adjustments are made with an immersion heater, and observations are based on a multimeter thermistor probe (Voltcraft VC870, Conrad Electronic SE, Germany). Upon increasing the temperature in the tank, it can be assumed that the temperature in the MMR will change rapidly in comparison to the surrounding water, due to the specific heat capacity of air being four times lower than that of water~\cite{stephan2019vdi}. The set temperatures are kept steady throughout the individual measurement process, following a protocol that used $20$ averaged frames per data point (sequence cycle: $t_\text{tx}=200$\,ms, $t_\text{rx}=3000$\,ms).

%%%%%%%%%%%%%%%%%%%%%%%%%%%%%%%%%%%%%%%%%%%%%%%%%%%%%%%%%%%%
\section{Results}
\subsection{Static measurements}
\label{sec:res:static}
The increased pressure on the MMR sensor impresses the membrane, reducing the distance between the magnets and thus increasing the measured oscillation frequency. Optical evaluation of the edge-to-edge distance between the magnets~$d_\text{e}$ (see Figure~\ref{fig:MMR-drawing}~(b)) and the static pressure levels are shown in Figure~\ref{fig:res-dist-and-static}~(a) and~(b), respectively. These results illustrate the relationship between magnet separation, natural frequency, and pressure. The error of the distances is larger for MMR~B, due to a different camera position and image resolution. The amount of plotted measurements of MMR~B in Figure~\ref{fig:res-dist-and-static}~(a) is reduced to improve visibility below \SI{168}{\Hz}. A non-linear trend of the natural frequency is observed, intensifying at small distances, which is in agreement with results in~\cite{knopp_empirical_2024}.  
Sensitivity of the $0.8$~mm membrane sensors can be approximated with $0.07~\text{Hz mbar}^{-1}$ and $0.06~\text{Hz mbar}^{-1}$ for MMR~A and B, respectively, in the region below $100$~mbar. Here, the standard deviation ($n=20$ each) in frequency of MMR~A is $0.015$~Hz ($0.25$~mbar). Over the total measurement range, the standard deviation for MMR~A and B average to $0.2$~Hz and $0.4$~Hz, respectively. Sensitivity increases with decreasing distance at higher pressure, due to the non-linearly increasing slope of the frequency.

\begin{figure}[t!]
    \centering
    \includegraphics[width=1.0\linewidth]{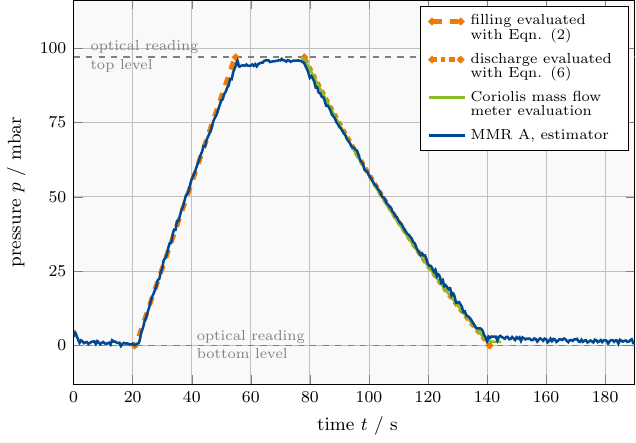}
    \caption{\textbf{Dynamic MMR measurement results.} Water is filled and discharged from the column, with MMR measurements taken at a 2~Hz frame rate. The natural frequency is evaluated for each time point (no averaging) and converted to pressure using an estimator based on a model fit of the static measurement. Theoretical hydrostatic pressure and measured hydrostatic pressure using a mass flow meter are provided for reference.}
    \label{fig:res-dynamic}
\end{figure}

\subsection{Dynamic measurements}
\label{sec:res:dynamic}
The temporal performance of MMR~A is tested with dynamic experiments at $2$~Hz frame rate, which is selected for its higher sensitivity. In order to map frequency to pressure, the inverse of a least squares estimator based on previous static experiments is employed. In Figure~\ref{fig:res-dynamic}, the measurement of MMR~A (blue) is compared with the hydrostatic pressure in the column, calculated using Equations~(\ref{eq:Hydrostatic}), (\ref{eq:Const-flow}), and~(\ref{eq:Toricelli}) (orange) with the measured times for inlet and outlet. The constant liquid levels are obtained optically, with the final level in the column determined to be $h = (99 \pm 1)$~cm, corresponding to a hydrostatic pressure of $p(h) = (96.9 \pm 0.98)$~mbar. Furthermore, the outlet is measured using the mass flow meter (green), evaluated and included into the plot.
Figure~\ref{fig:res-dynamic} confirms that Equation~(\ref{eq:Toricelli}) is consistent with the measurement obtained using the mass flow meter.  The overall pressure measurement is also in agreement with theory and measured pressure, with the highest deviation on the plateaus. Maximum errors of $3.7$~mbar and $4.4$~mbar occurred on the high and low plateaus, respectively, with an overall standard deviation of $1.57$~mbar.
The average zero-pressure levels at the beginning and end align within manual operation tolerances, and the membrane material exhibits no significant hysteresis, returning to its equilibrium state at $p_0$ during this experiment. At stationary levels, a noisy frequency component equalling the sequence repetition frequency of $2$\;Hz can be observed.

\begin{figure}[t!]
    \centering
    \includegraphics[width=1.0\linewidth]{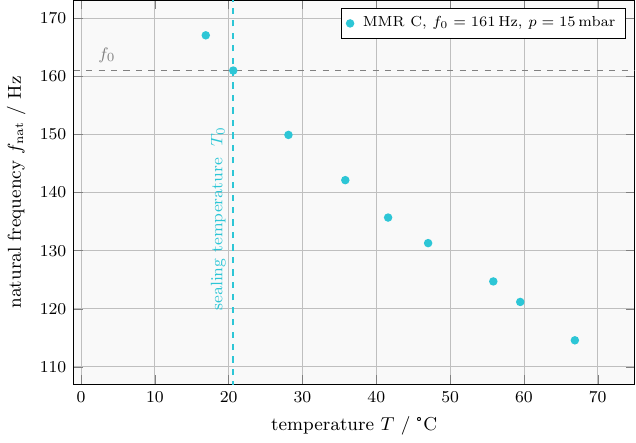}
    \caption{\textbf{Temperature measurement results.} The MMR is sealed at ambient temperature $T_0$ and an increase in the ambient water temperature (at constant pressure $p=15$\,mbar) causes the pressure inside the sensor to increase, separating the magnets and resulting in a reduction in frequency.}
    \label{fig:res-temperature}
\end{figure}

\subsection{Temperature dependence}
\label{sec:res:temp}
Figure~\ref{fig:res-temperature} illustrates the temperature dependence, affirming the trend that the natural frequency decreases with increasing temperature. As the temperature rises, the pressure within the confined space of the MMR increases due to changes in the density of the encapsulated air. This is supported by the fact that the air density changes by 15~\% within the shown temperature range~\cite{stephan2019vdi}. Optical observations also confirmed this behaviour, with the membrane curving away from the casing. Consequently, the net pressure on the membrane changes, increasing the distance between the magnets. It is anticipated that the temperature effect will counteract the pressure effect. A linear regression indicates that the sensor is sensitive to temperature changes with $0.98$\;Hz $^\circ$C$^{-1}$ and a coefficient of determination $r^2$ of $0.985$ for the measured interval. 
%%%%%%%%%%%%%%%%%%%%%%%%%%%%%%%%%%%%%%%%%%%%%%%%%%%%%%%
\section{Discussion}
The choice of a $0.8$~mm membrane thickness proves to be durable, with the weakest point of the sensor being the connection between membrane and thread. At this point, the material experiences the highest stress and deformation. 
Averaging of static measurements shows that the MMR pressure sensor can achieve a standard deviation of $0.25$~mbar in the range $0$ to $100$~mbar. The standard deviation is higher for dynamic measurements at $2$~Hz without averaging at $1.57$~mbar, but remains below 2~\%. The non-linear behaviour of the sensor in Figure~\ref{fig:res-dist-and-static} is expected from the non-linear characteristic of $\Delta d$ (see Figure~\ref{fig:force}) and from the dipole model assumption in Equation~(\ref{eq:fnat})~\cite{gleich_miniture_2023, knopp_empirical_2024}. However, the results show that it can be calibrated successfully based on static measurements. Higher averaging or a sufficiently accurate physical model may improve precision and accuracy. MMR pressure sensors are below specification of commercially available sensors such as piezoresistive or capacitive pressure sensors, but with the great advantage of being wireless and passive. There is considerable scope for improvement and optimisation in all aspects of the presented sensor, including its manufacturing process, the mechanism by which pressure is coupled to frequency (housing design) via a single membrane, and signal detection equipment. \par
As shown in Figure~\ref{fig:res-temperature}, the MMR has a significant cross sensitivity to temperature. While temperature sensitivity is not a design parameter in this study, as the primary focus is on assessing pressure sensitivity, future optimisations will aim to reduce the volume of air within the MMR to mitigate its influence on temperature dependence. Furthermore, it is important to acknowledge that the relationship between frequency and temperature may not be linear. 
It is therefore not possible to simultaneously quantify temperature and pressure with this MMR design. However, if one of the two parameters is known, calibration can be used to compensate for the frequency shift due to the other parameter. Moreover, modern bioprocesses, as a potential future application for MMRs, are aimed to maintain constant temperatures and uniform temperature distribution within bioreactor vessels~\cite{simutis2015bioreactor, kumar2019temperature}.\par
%%% 
The absolute range of the proposed MMR sensor is constrained by the maximum stretch distance~$\Delta d$, which can be modified through the selection of membrane durability and thickness, in consideration of sensitivity. With thicker membranes, the sensitivity is lower but the pressure range is wider. A reduction in the initial distance of the sensor enhances the sensitivity due to the increased slope, which in turn restricts the range. The thickness of the membrane and initial magnet distance should therefore be tailored to the specific requirements of the application. \par
%%% 
The presence of electromagnetic noise and distorted supply currents in process engineering do not impair the sensitivity at the specified distance. Detection limitations of passive resonant sensors are based on the steep drop in \ac{SNR} due to distance, which is further deteriorated by conductive fluids or vessels because of eddy currents. The experiments presented in this study are conducted within a transparent and non-conductive column. However, industrial vessels are generally manufactured from optically inaccessible and conductive materials (in process engineering). In this instance, a promising result is observed in an experiment where the inductive measurement setup is tested with an non-transparent, conductive column made from 1~mm thick aluminium, which results in a reduction of the \ac{SNR} by approximately a factor of two. \par
On the other side, the emitted signal of the sensor has an upper limit determined by physical parameters such as frequency (induction), deflection angle, and magnetic moment (sensor size), even if excitation is achieved at larger distances using high and pulsed currents. Increasing the \ac{SNR} can be achieved with ultra low noise amplifiers, dedicated receive coils, and stronger magnetic materials (N52 or higher). Smaller MMRs allow a higher frame rate due to their higher frequencies, enabling faster data acquisition and more averaging. \par
%%% 
Accuracy of calculating $f_\text{nat}$ depends on the model input parameters like windowing and iterations as well as on convergence of the model~\cite{knopp_empirical_2024}. The deviation between the modelled pressure and estimator MMR pressure in Figure~\ref{fig:res-dynamic}, depends on the fitting and static pressure levels of Figure~\ref{fig:res-dist-and-static}. Results can be further improved by using closer spaced calibration values and more averages. \par

The technology is well-suited to harsh environments, which are commonly found in industrial process engineering. Additionally, the measurement technology is capable of operating through acrylic glass and water. However, further investigations are required to assess the penetration capabilities of reactor materials and media.  
The design of the MMR housing is fundamental to address the central challenges laid out in this work. These include the sensitivity to pressure, the cross-sensitivity to temperature, and buoyancy. In the future, MMRs will be further developed to enable flow-following capabilities. To achieve this, the density of the MMR must be adjusted to match that of the surrounding medium. Since the MMRs tested in this study exhibit positive buoyancy in water, the volume of air within the MMRs needs to be reduced or replaced with alternative media, such as gel. 
%%%%%%%%%%%%%%%%%%%%%%%%%%%%%%%%%%%%%%%%%%%%%%%%%%%%%%%
\section{Conclusion}
An MMR pressure sensor has been successfully developed and tested. The sensor employs a $0.8$~mm 3D printed flexible membrane. The adaptability of the additive manufacturing process allows for the modification of membrane geometry and material to align with the desired pressure sensitivity, pressure range, and specific application. The MMR design demonstrates a minimum sensitivity of $0.06~\text{Hz mbar}^{-1}$, with sensitivity increasing at higher pressures, making it suitable for process engineering applications. The relationship between frequency and pressure is non-linear and requires calibration, for example using a model fit of static measurements. Additionally, the sensor’s maximum pressure of 600~mbar, approximately equivalent to a hydrostatic pressure of 6~m, accommodates a wide range of reactors commonly used in the field. The analysis of the dynamic measurements reveals an average accuracy of 1.57~mbar. Averaging the real-time readout at $2$~Hz can enhance the accuracy of the measurement, contingent on the desired dynamic constraints. The MMR exhibits a notable cross-sensitivity to temperature with 0.98 Hz $^\circ$C$^{-1}$. \par
A significant advantage of MMR pressure sensors over Lagrangian sensor particles is their ability to be localised in terms of position and orientation within the reactor~\cite{gleich_miniture_2023}, wirelessly and through opaque fluids such as emulsions, flows with high cell or bubble densities, and optically inaccessible materials. Additionally, the MMRs are passive, negating the need for a power supply and thereby simplifying the device's design, which also enables cost-effective manufacturing. This measurement principle further eliminates the risk of battery-derived hazardous substances, enhancing their suitability for industrial bioreactors.
%%%%%%%%%%%%%%%%%%%%%%%%%%%%%%%%%%%%%%%%%%%%%%%%%%%%%%%
\section*{Contributions~\&~acknowledgments}
T.M. printed the 3D membranes and supervised the force experiments.
T.M. and F.M. manufactured the MMR sensors.
T.M., F.K., J.F., and F.M. conducted experiments 3.1 and 3.2.
F.M. and J.F. conducted experiment 3.3.
M.S., T.K., T.M., M.M., F.K., J.F., and F.M. contributed to the conceptualisation and theory. 
M.S. and T.K. supervised the project.
T.M. and F.M. wrote the original draft with support from F.K., J.F., and M.M.
All authors reviewed the final manuscript.
The authors would like to thank Furkan Kahraman for his help in building the separable Helmholtz coil.
Publishing fees supported by Funding Program Open Access Publishing of Hamburg University of Technology (TUHH).
This project is funded by the Deutsche Forschungsgemeinschaft (DFG, German Research Foundation) – SFB 1615 – 503850735.
%%%%%%%%%%%%%%%%%%%%%%%%%%%%%%%%%%%%%%%%%%%%%%%%%%%%%%%
\section*{Data availability statement}
The datasets generated for this study can be found in the TUHH Open Research (TORE) \href{https://doi.org/10.15480/882.14543}{\textsc{DOI}: 10.15480/882.14543}.
%%%%%%%%%%%%%%%%%%%%%%%%%%%%%%%%%%%%%%%%%%%%%%%%%%%%%%%

%%%%%%%%%%%%%%%%%%%%%%%%%%%%%%%%%%%%%%%%%%%%%%%%%%%%%%%
\printbibliography%

\end{document}